\documentclass[twocolumn,showpacs,amsmath,amssymb,prl]{revtex4}

\pdfoutput=1

\usepackage{graphicx}
\usepackage{dcolumn}
\usepackage{bm}
\usepackage{epsf}
\usepackage{amsmath}
\usepackage{amssymb}
\usepackage{color}

\newcommand{\la}{\left\langle}
\newcommand{\ra}{\right\rangle}
\newcommand{\pd}{\partial}

\newcommand{\td}{\mathrm{d}}

\newcommand{\e}[1]{\exp{\left(#1\right)}}

\newcommand{\bla}{bla\\bla\\bla\\bla\\bla}
\newcommand{\PR}{Phys. Rev. }

\newcommand{\PRE}{Phys. Rev. E }
\newcommand{\PRL}{Phys. Rev. Lett. }
\newcommand{\EPL}{EPL }

\begin{document}

\title{Information free energy for nonequilibrium states} 

\author{Sebastian Deffner$^{1,3}$ and Eric Lutz$^{2,3}$ }

\affiliation{$^1$Department of Chemistry and Biochemistry and Institute for Physical Science and Technology, University of Maryland, 
College Park, Maryland 20742, USA\\$^2$Dahlem Center for Complex Quantum Systems, FU Berlin, D-14195 Berlin, Germany\\
$^3$Department of Physics, University of Augsburg, D-86135 Augsburg, Germany}

\date{\today}

\begin{abstract}
We introduce an information free energy for thermodynamic systems driven by
external time-dependent parameters. We show that the latter is a
nonequilibrium state function and that it is a natural generalization of
the usual equilibrium and steady state free energies. We discuss its importance for the
nonequilibrium maximum work theorem
 and the Jarzynski relation in the presence of  feedback control.  We further show that it is a nonequilibrium Lyapunov function.
\end{abstract}


\maketitle

Thermodynamics is a theory of  processes and states. Thermodynamic functions are accordingly  divided into process-dependent quantities, like work and heat, and state variables that do not depend on specific transformations, but only on the current state of a system \cite{cal85}. Contrary to work and heat, thermodynamic state functions are only defined at equilibrium. Two important examples of equilibrium variables are the free energy $F$ and the entropy $S$, which are both related to work or  heat for quasistatic processes: the isothermal change of free energy is equal to the mean work,  $\Delta F= \la W\ra$, while the change of entropy is given by the mean heat divided by temperature,   $\Delta S = \la Q\ra/T$. A central question is how to extend these well-established concepts to arbitrary processes and  nonequilibrium states. The Shannon information entropy of a probability distribution $p(x)$ is defined as $\mathcal{H} =- \int \td x \,p(x) \ln p(x) $ \cite{sha48}. Remarkably, for equilibrium transformations  both entropies coincide, $\Delta \mathcal{H}= \Delta S$, and the variation of information entropy  is thus also determined by the mean heat \cite{jay57} (we here set the Boltzmann constant to unity). This identity illustrates the intimate connection  between thermodynamics and information theory. The Shannon entropy $\mathcal{H}$, being defined for any probability distribution,  is commonly used  as a nonequilibrium  entropy  \cite{set06,gra08,sei05,ge10}. However, its precise relationship to the mean heat out of  equilibrium seems to be unclear. Moreover, a proper generalization of the free energy valid far from equilibrium appears to be lacking.

In the  present article, we introduce an information based free energy $\mathcal{F}$ that is a  nonequilibrium state function. We show that it reduces to the usual free energy for equilibrium states, and to the free energy of Hatano and Sasa for nonequilibrium steady states \cite{hat01}. We, moreover, highlight its essential role in the formulation of the  nonequilibrium maximum work theorem \cite{cal85},  and in the generalized  work relation under nonequilibrium feedback control \cite{sag10}. In addition, we establish that  it decays monotonously during relaxation processes towards equilibrium or nonequilibrium steady states, and that it is, hence,  a proper nonequilibrium Lyapunov function \cite{mae11}.

\paragraph{Information free energy.} In the following, we consider the paradigmatic model of an overdamped Brownian particle moving in a time-dependent potential $V(x,\alpha)$, where $\alpha=\alpha_t$ is an external control parameter. The time evolution of the particle is described by the Smoluchowski equation for the probability density $p(x,t)$ \cite{ris89},
\begin{equation}
\label{q01}
\begin{split}
\pd_t\, p(x,\alpha)&= L_\alpha\, p(x,\alpha)\\
\mathrm{with}\hspace{1em}L_\alpha &=\frac{1}{\gamma}\,\pd_x\left[ V'(x,\alpha)+\frac{1}{\beta}\,\pd_x\right]\,,
\end{split}
\end{equation}
where $\gamma$ is the damping coefficient and $\beta$  the inverse temperature. For  a particular value of the parameter $\alpha_t$, the stationary  state verifies $\pd_t\, p(x,\alpha_t) = L_\alpha\, p(x,\alpha_t)=0$. We begin by treating  the case of  equilibrium steady states and investigate the connection between  information entropy and heat in the quasistatic limit. For  quasistatic driving $ \dot \alpha_t \rightarrow 0$, we have $\pd_t\, p(x,\alpha_t) = \int \td t\, \dot \alpha_t \, \pd_\alpha \, p(x,\alpha_t) \rightarrow 0$. We note, however, that for a general (nonequilibrium) distribution, $L_\alpha\,  p(x,\alpha_t)\neq 0$. Let us, now, consider the difference between the change of the Shannon entropy  and the mean heat,
\begin{equation}
\label{q02}
\la\Sigma\ra=\Delta \mathcal{H} -\beta\la Q\ra.
\end{equation}
The quantity $\la\Sigma\ra$ may be regarded as an information entropy production \cite{sei05}. The average heat exchanged with the bath is given by the usual definition \cite{sek10}, 
\begin{equation}
\label{q03}
\beta\la Q\ra=\int_0^\tau \td t\,\int\td x\,\pd_t\, p(x,\alpha_t )\,\beta V(x,\alpha_t).
\end{equation}
At the same time, the variation of the Shannon entropy can be written in the form,
\begin{equation}
\label{q04}
\Delta \mathcal{H}=-\int_0^\tau\td t\,\int\td x\,\pd_t\, p(x,\alpha_t)\,\ln{p(x,\alpha_t)}\,.
\end{equation}
Combining Eqs.~\eqref{q03} and \eqref{q04},  and using the Smoluchowski  equation \eqref{q01}, we obtain after integrations by parts,
\begin{equation}
\label{q05}
\la\Sigma\ra=\int_0^\tau \td t\int\td x\,\frac{\left(p'(x,\alpha_t)+\beta V'(x,\alpha_t)\,p(x,\alpha_t)\right)^2}{\beta\gamma\,p(x,\alpha_t)}\,.
\end{equation}
The above expression shows that $\Delta \mathcal{H} = \beta\la Q\ra$ only if the current $j(x,\alpha_t)=p'(x,\alpha_t)+\beta V'(x,\alpha_t)\,p(x,\alpha_t) = 0$, i.e. if the system is in a stationary state, $L_\alpha\, p(x,\alpha_t)=0$, at all times. The latter condition is fulfilled for  $\dot \alpha_t \rightarrow 0$ \textit{and}  initial stationary states $L_{\alpha_0}\, p(x,\alpha_0) = 0$. We can, thus, conclude that for arbitrary nonequilibrium initial states, the change of  information entropy is not given by  the mean heat, even for quasistatic driving. The latter can be understood by noticing that an initial nonequilibrium state will relax spontaneously to the equilibrium state, and the relaxation time is in general different from the driving time. Thus, relaxation needs not be slow, even if the driving is. As a result, the mean information entropy production is strictly positive in that instance, $\la \Sigma \ra>0$. 

Equation \eqref{q05} has important consequences for the maximal amount of work that can be extracted from a given system. Using the first law of thermodynamics, $\Delta U =U(\alpha_\tau)- U(\alpha_0)= \la W\ra + \la Q\ra$, with $U(\alpha)= \int\td x\,p(x,\alpha)\, V(x,\alpha)$, we find,
\begin{equation}
\label{q06}
\begin{split}
\la\Sigma\ra= \Delta \mathcal{H}-\beta\la Q\ra 
& =\Delta \mathcal{H}-\beta \Delta U+\beta \la W\ra\\
&=\beta \la W\ra-\beta \Delta \mathcal{F}\geq0\ ,
\end{split}
\end{equation}
where we have introduced the information free energy,
\begin{equation}
\label{q07}
\begin{split}
\beta\,\mathcal{F}(\alpha_t)&=\beta\, U(\alpha_t)-\mathcal{H}(\alpha_t)\\
&=D\left(p(x,\alpha_t)||p_\mathrm{eq}(x,\alpha_t)\right)+\beta F(\alpha_t)\ .
\end{split}
\end{equation}
Here $D\left(p(x,\alpha)||p_\mathrm{eq}(x,\alpha)\right) = \int \td x\, p(x,\alpha) \ln p(x,\alpha) - \int \td x\,p(x,\alpha) \ln p_\mathrm{eq}(x,\alpha)$ is the relative (Kullback-Leibler) entropy \cite{kul51,kul78} between $p(x,\alpha)$ and the  equilibrium distribution $p_\mathrm{eq}(x,\alpha)=\exp{(-\beta V(x,\alpha))}/Z(\alpha)$ corresponding to the same  parameter $\alpha$.  The equilibrium free energy reads $F(\alpha) = -(1/\beta) \ln Z(\alpha)$, where $Z(\alpha)$ is the partition function. The information free energy \eqref{q07} is a nonequilibrium state function, since for a cyclic process between arbitrary nonequilibrium states, 
\begin{equation}
\label{q08}
\oint \td\mathcal{F}= \mathcal{F}(\alpha_0)-\mathcal{F}(\alpha_0)= 0, \quad \alpha_0=\alpha_\tau.
\end{equation}
The  information free energy $ \mathcal{F}$ is strictly speaking a relative quantity that depends both on the actual state of the system, $p(x,\alpha_t)$ and the reference equilibrium state, $p_\mathrm{eq}(x,\alpha_t)$. However, by considering the latter to be fixed (for given potential and temperature), $ \mathcal{F}$ can be interpreted as depending on the arbitrary state $p(x,\alpha_t)$ alone.
Equation \eqref{q06} is an expression of the maximum work theorem, extended to arbitrary nonequilibrium states \cite{tak10,tak10a}: the maximum work, $-\la W\ra$, that can be extracted from a system is always smaller than the change of information free energy, $-\Delta \mathcal{F}$. The  bound is tight  only for stationary states in the quasistatic limit. 

The  information free energy \eqref{q07} is defined for arbitrary nonequilibrium states.  For transitions between two equilibrium states, $p_\mathrm{eq}(x,\alpha')$ and $p_\mathrm{eq}(x,\alpha)$, the relative entropy simplifies to $D\left(p_\mathrm{eq}(x,\alpha')||p_\mathrm{eq}(x,\alpha)\right)=  F(\alpha') - F(\alpha)$ \cite{don87}, and  $\mathcal{F}(\alpha')$ reduces to the equilibrium free energy $ F(\alpha')$. On the other hand, for transitions between  two nonequilibrium steady states,  $p_\mathrm{ss}(x,\alpha')$ and $p_\mathrm{ss}(x,\alpha)$, the information free energy reduces to the steady state free energy  defined by Hatano and Sasa \cite{hat01}.

We may  indeed repeat the above analysis for nonequilibrium steady states induced by either a driven potential \cite{tre04} or  a nonconservative force $f_t$ \cite{bli07,gom09} by considering, in Eq.~\eqref{q01}, the operator
\begin{equation}
\label{q09}
L_\alpha=\frac{1}{\gamma}\,\pd_x\left[ V'(x,\alpha)+f_t+\frac{1}{\beta}\,\pd_x\right].
\end{equation}
In this case, following Hatano and Sasa \cite{hat01} (see also  Ref.~\cite{oon98}), we divide the total heat in house-keeping heat (exchanged in a given nonequilibrium steady state) and excess heat (exchanged during a change of steady states),   $Q_\mathrm{tot}=Q_\mathrm{hk}+Q_\mathrm{ex}$.  For  equilibrium steady states, we simply have $Q_\mathrm{tot}=Q_\mathrm{ex}$.  The excess heat reads \cite{hat01},
\begin{equation}
\label{q10}
\beta\la Q_\mathrm{ex}\ra=\int_0^\tau \td t\,\int\td x \,p(x,\alpha_t)\, \dot x\,\pd_x \varphi(x,\alpha_t)\ ,
\end{equation}
where ${\varphi(x,\alpha_t) = -\ln p_\mathrm{ss}(x,\alpha_t)}$ and $p_\mathrm{ss}(x,\alpha_t)$ is the nonequilibrium steady state,  $L_\alpha\,p_\mathrm{ss}(x,\alpha)=0$. The difference between the information entropy change and the mean excess heat, $\la \Sigma_\mathrm{ex}\ra=\Delta\mathcal{H}-\beta \la Q_\mathrm{ex}\ra$, can then  be determined as done previously. We find
\begin{equation}
\label{q11}
\la\Sigma_\mathrm{ex}\ra=\int_0^\tau \td t\int\td x\,\frac{j(x,\alpha_t)^2}{\beta\gamma\,p(x,\alpha_t)}\,.
\end{equation}
Accordingly,  $\Delta \mathcal{H}= \beta \la Q_\mathrm{ex}\ra$ only if $j(x,\alpha_t) = p'(x,\alpha_t)+\beta\left( V'(x,\alpha_t)+f_t\right)\,p(x,\alpha_t) = 0$ at all times, that is only for quasistatic processes starting in initial stationary states. 
Introducing the excess work as $\la W_\mathrm{ex}\ra= \Delta U - \la Q_\mathrm{ex}\ra$ and the information free energy \eqref{q07}, the maximal work theorem takes here the form,
\begin{equation}
\label{q12}
\la W_\mathrm{ex}\ra\geq  \Delta \mathcal{F}\,.
\end{equation}
Equation \eqref{q12} generalizes the inequality obtained by Hatano and Sasa \cite{hat01} to arbitrary initial nonequilibrium states. Again, the bound is  tight only  for initial stationary states and quasistatic driving. 

\paragraph{Nonequilibrium feedback control.} The information free energy \eqref{q07} finds also application in the context of  feedback control. The Jarzynski equality $\la \exp(-\beta W)\ra = \exp(-\beta \Delta F)$ \cite{jar97} has been extended by Sagawa and Ueda to account for external information gain \cite{sag10}. The latter result is however restricted to initial equilibrium states, and a generalization to arbitrary states is therefore of interest \cite{abr12}. Following Ref.~\cite{sag08}, we consider the situation where 
the parameterization of the control parameter  $\alpha$ is updated during the process by feedback measurements. We denote by $X=\{x_t\}^{\tau}_0$ a trajectory of the system that starts at $t=0$ and ends at $t=\tau$, and  suppose that  a measurement on $X$ is performed at time $t$. Let $x_t$ be the space point of the system, $P[x_t]$ its probability, and $y$ the measurement outcome---we thus have $\alpha=\alpha(t,y)$. We further assume that the measurement can involve an error which is described by the conditional probability, $P[y|x_t]$, to measure $y$ while the system is at position $x_t$. The probability of the measurement outcome $y$ is then,
\begin{equation}
\label{q13}
P[y]=\int\td x_t\,P[y|x_t]P[x_t]\,. 
\end{equation}
The information obtained from the measurement is characterized by the mutual information \cite{sag10},
\begin{equation}
\label{q14}
 \la I\ra=\int\td x_t \int\td y\,P[y|x_t]P[x_t]I[x_t,y]\,,
\end{equation}
where $I[x_t,|y]$ is given by
\begin{equation}
\label{q15}
 I[x_t,y]=\ln{\left(\frac{P[y|x_t]}{P[y]}\right)}\,.
\end{equation}
Following the path integral approach elaborated in Refs.~\cite{che06,def11a}, we introduce the probability of a forward trajectory $P^F[X]$. The time reversed counterpart is accordingly  $P^R[X^\dagger]=\e{-\Sigma} P^F[X]$ \cite{che06,def11a}. As a result, \begin{equation}
\label{q16}
\begin{split}
& \la\e{-\Sigma-I} \ra=\\
&=\int {\cal D}X\int\td y\,P[y|x_t]\,P^F\left[X\right]\,\e{-\Sigma}\,\frac{P[y]}{P[y|x_t]}\\
&=\int {\cal D}X^\dagger\int\td y\,P^R\left[X^\dagger\right]\,P[y]=1\,,
\end{split}
\end{equation}
where we have used Eq.~\eqref{q15} and the fact  that $\mathcal{D}X=\mathcal{D}X^\dagger$. The last equality follows from normalization. Expression \eqref{q16} is a generalization of the Jarzysnki equality with  feedback to initial nonequilibrium states. 
An application of Jensen's inequality yields,
\begin{equation}
\label{q17}
 \la \Sigma\ra+\la I\ra\geq0\,.
\end{equation}
Making eventually use of the explicit expression of the entropy production in  a nonequilibrium steady state, $\Sigma = \Sigma_\mathrm{ex}=\int_0^\tau\td t\,\pd_\alpha \varphi\,\dot\alpha$, \cite{che06,def11a}, we obtain,
\begin{equation}
\label{q18}
\begin{split}
\la \Sigma_\mathrm{ex}\ra&=\la \int_0^\tau\td t\,\pd_\alpha \varphi\,\dot\alpha\ra\\
&=\la \Delta\varphi\ra-\int_0^\tau\td t\int\td x \,p(x,t)\,\dot x\,\pd_x\varphi\\
&=\Delta\mathcal{H}-\beta\la Q_\mathrm{ex}\ra=\beta\la W_\mathrm{ex}\ra-\beta\Delta\mathcal{F}.
\end{split}
\end{equation}
The generalization of the second law to nonequilibrium steady states with feedback takes therefore the form,
\begin{equation}
\label{q19}
 \beta\la W_\mathrm{ex}\ra\geq\beta\Delta\mathcal{F}-\la I\ra\,.
\end{equation}
A similar calculation for arbitrary nonequilibrium states leads to a general result of the same form for $\la W\ra$, showing that more work can be extracted from a system in the presence of feedback \cite{abr11}. The latter reduces to the special result derived by Sagawa and Ueda for equilibrium states when  $\mathcal{F}$ is replaced by $F$.

\paragraph{Nonequilibrium Lyapunov function.} Equilibrium states are stable and correspond to a minimum of the free energy \cite{cal85}. The relaxation of a nonequilibrium perturbation back to the equilibrium state is often characterized by the monotonic decay of a Lyapunov function. The question of  the existence of Lyapunov functions for the relaxation towards general nonequilibrium steady states has recently been raised \cite{mae11}. It is often believed that the free energy
difference  should be replaced by the relative entropy away from equilibrium. However, due to its oscillating behavior, the latter is not a  Lyapunov function \cite{mae11}. The central  point of the present paper is that the correct nonequilibrium generalization of the free energy is the information free energy \eqref{q07}. For purely relaxation processes  towards a steady state ($\la W_\mathrm{ex}\ra=0$), its time derivative is simply, 
\begin{equation}
\label{q21}
\beta\,\td_t \mathcal{F}(\alpha)=-\td_t\la \Sigma_\mathrm{ex}\ra \leq 0.
\end{equation}
The information free energy hence  decays monotonously to its minimal value at all times---and not only asymptotically (see Ref.~\cite{mae11}). It is therefore a true Lyapunov function for equilibrium and nonequilibrium steady states.

\paragraph{Examples.} To illustrate the results \eqref{q06}, \eqref{q12} and \eqref{q21} for equilibrium and nonequilibrium steady states, we first consider a Brownian particle in a harmonic trap with time-dependent frequency, $V(x,\alpha_t)=\alpha_t\,x^2/2$. This system has been studied experimentally to verify the fluctuation theorem \cite{car04}. For concreteness,  we choose the initial {(nonequilibrium)} distribution $p(x,0)$ to be Gaussian. Due  to the linearity of the Smoluchowski equation \eqref{q01}, its time-dependent  solution $p(x,t)$ is  also Gaussian.
Its  mean  $\mu_t$ and variance $\sigma_t$ satisfy the equations,
\begin{subequations}
\label{q22}
\begin{eqnarray}
 \dot \mu_t&=&\left(\frac{2 \dot \sigma_t}{\sigma_t}-\frac{2}{\beta \gamma}\frac{1}{\sigma_t^2}+\frac{\alpha_t}{\gamma}\right)\,\mu_t,\\
 \dot \sigma_t&=&\frac{1}{\beta\gamma}\frac{1}{\sigma_t}-\frac{\alpha_t}{\gamma}\,\sigma_t.
\end{eqnarray}
\end{subequations}
The  corresponding time-dependent solutions are 
\begin{subequations}
\label{q23}
\begin{eqnarray}
\mu_t&=&\mu_0\,\e{-\frac{1}{\gamma}\int_0^t\td s\,\alpha_s}\,,\\
 \sigma_t^2&=&\sigma_0^2\,\e{-\frac{2}{\gamma}\int_0^t\td s\, \alpha_s} \nonumber\\
&+& \frac{2}{\beta\gamma}\int\limits_0^t\td s\,\e{-\frac{2}{\gamma}\int_0^{s}\td s'\, \alpha_{s'}}.
\end{eqnarray}
\end{subequations}
The mean heat  \eqref{q03} exchanged with the bath follows as,
\begin{equation}
 \label{q24}
\beta\la Q\ra=\beta\int_0^\tau \td t\,\alpha_t\,\left(\mu_t\,\dot\mu_t+\sigma_t\,\dot\sigma_t\right),
\end{equation}
and the variation of information entropy \eqref{q04} reads
\begin{equation}
 \label{q25}
 \Delta\mathcal{H}=-\frac{1}{2}\,\ln{\frac{\sigma_\tau^2}{\sigma_0^2}}.
\end{equation}
On the other hand,  the mean work  is
\begin{equation}
\label{q26}
\begin{split}
 \la W\ra&=\int_0^\tau \td t\int\td x\,p(x,t)\,\pd_\alpha V\,\dot \alpha_t\\
&= \frac{1}{2}\int_0^\tau \td t\,\dot \alpha_t\,\left(\mu_t^2+\sigma_t^2\right)\, ,
\end{split}
\end{equation}
and the information  free energy difference is  given by
\begin{equation}
\label{q27}
\Delta \mathcal{F}=\frac{\alpha_\tau}{2}\,\left(\mu_\tau^2+\sigma_\tau^2\right)-\frac{\alpha_0}{2}\,\left(\mu_0^2+\sigma_0^2\right)-\frac{1}{2\beta}\,\ln{\frac{\sigma_\tau^2}{\sigma_0^2}}\,.
\end{equation}
We notice that these two last expressions usually differ. In the quasistatic limit, $\dot \alpha_\tau \simeq (\alpha_\tau -\alpha_0)/\tau \rightarrow 0$, the first two terms on the right-hand side of Eq.~\eqref{q27} reduce to the mean work \eqref{q26}.  However, the last term in Eq.~\eqref{q27} does not depend on the driving rate and only vanishes if 
$\sigma_\tau^2=\sigma_0^2$, i.e. when the system already starts in the equilibrium state $p(x,0)=p_\mathrm{eq}(x,0)$.
The time derivative of the information free energy \eqref{q07} further reads
\begin{equation}
\label{q28}
\td_t\mathcal{F}(\alpha_t)=-\frac{1}{\beta^2\gamma}\int\td x\,p(x,\alpha_t)\,\left(\beta\,\alpha_t\, x-\frac{x-\mu_t}{\sigma_t^2}\right)^2,
\end{equation}
which is clearly strictly negative at all times.

\begin{figure}
\centering
 \includegraphics[width=0.48\textwidth]{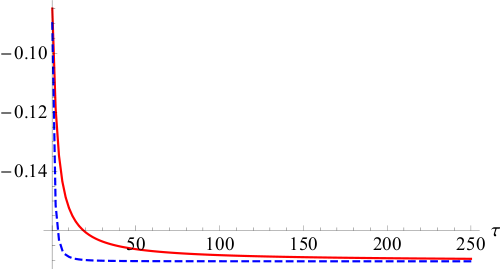}
\caption{\label{fig1}(color online) Average excess work $\la W_\mathrm{ex}\ra$ (red, solid) and information free energy difference $\Delta\mathcal{F}$ (blue, dashed) as a function of the driving time $\tau$ for a particle  in the tilted periodic potential \eqref{q27},  starting in the stationary state   \eqref{q29}. Parameters are $\gamma=1$, $\beta=25$, $k=1$, $h_0=0.2$, and $h_1=0.6$.}
\end{figure}

\begin{figure}
\centering
 \includegraphics[width=0.48\textwidth]{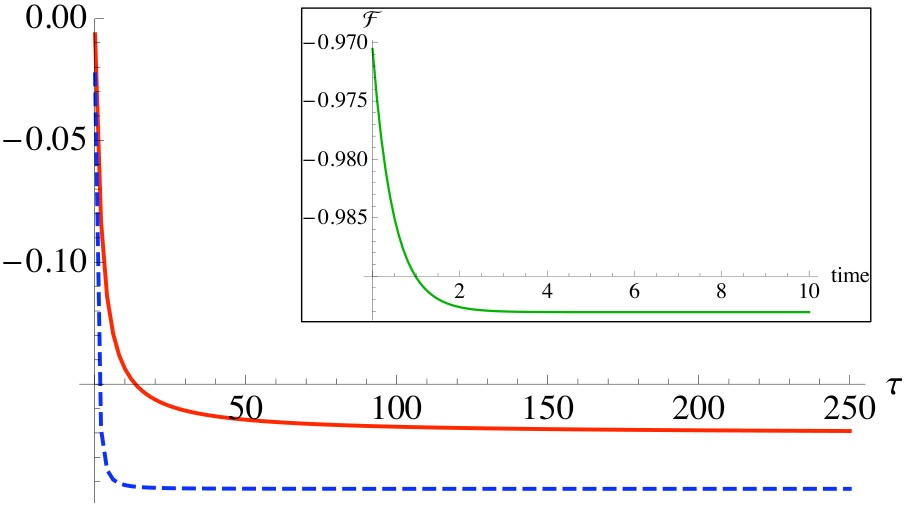}
\caption{\label{fig2}(color online) Average excess work $\la W_\mathrm{ex}\ra$ (red, solid) and information free energy difference $\Delta\mathcal{F}$ (blue, dashed)  as a function of the driving time $\tau$ for a particle in the tilted periodic potential  \eqref{q27}, starting in a nonstationary Gaussian state (see text). Parameters are $\sigma_0=0.25$, $\gamma=1$, $\beta=25$, $k=1$, $h_0=0.2$, and $h_1=0.6$. The inset shows the monotonic decay of the information free energy $\mathcal{F}$ as a function of time for the same initial state, but no external driving.}
\end{figure}

To examine the case of  a nonequilibrium steady state, we next consider a particle in a  tilted periodic potential, $V(x,t)=-k \cos x - x h(t)$, with $h(t) = h_0- (h_0-h_1)t/\tau$. The stationary solution of Eq.~\eqref{q01}, with periodic boundary conditions and $V'(x) = V'(x+2\pi)$, is \cite{rei02},
\begin{equation}
\label{q29}
p_\mathrm{s}(x)=\frac{\e{-\beta V(x,t)}}{N_t}\,\int_x^{x+2\pi}\td x\,\e{\beta V(x,t)}, 
\end{equation}
where $N_t$ is the normalization constant.  Figure~\ref{fig1} shows the mean excess work $\la W_\mathrm{ex}\ra$ and the information free energy difference $\Delta\mathcal{F}$ as a function of the duration of the process $\tau$,  when the particle is initially in the  stationary state \eqref{q29}. As expected, in the quasistatic limit, $\tau \rightarrow \infty$,  $\la W_\mathrm{ex}\ra $ approaches the lower bound $\Delta\mathcal{F}$. By contrast, in Fig.~\ref{fig2} the particle starts initially in a nonstationary Gaussian state $p(x,0)=(2\pi\,\sigma_0^2)^{(-1/2)}\,\exp(-x^2/(2\sigma_0^2))$. We observe  that even for  long driving times $\tau$, the mean work never reaches $\Delta\mathcal{F}$, due to the nonstatic relaxation of the initial state. In the absence of driving, the inset shows that the Lyapunov function $\mathcal{F}$ decays monotonously.

\paragraph{Conclusions.} Two state functions can be defined for arbitrary nonequilibrium states beyond linear response: the information entropy $\mathcal{H}$ and the information free energy $\mathcal{F}$. Both can be directly related to measurable thermodynamic quantities like (excess) work and heat  for  quasistatic processes that start in equilibrium (nonequilibrium) steady states. For arbitrary transformations, they seem to loose their thermodynamic interpretation. This is  especially true for quasistatic transformations that start in nonstationary states. However, they nevertheless provide useful bounds for the mean (excess) work and heat exchanged. In particular, the information free energy is a natural generalization of the equilibrium  and the steady state free energy to which it reduces in the appropriate limits. We have shown that it naturally appears in the nonequilibrium extensions of the maximum work theorem with or without feedback. Moreover, since it decays monotonously  to steady states at all times, it is a nonequilibrium Lyapunov function. Our results may also be extended to other types of dynamics \cite{ge10,def11,esp11}.

\acknowledgements{We gratefully acknowledge fruitful discussions with Christopher Jarzynski. This work was supported by the Emmy Noether Program of the DFG (contract No LU1382/1-1) and the cluster of excellence Nanosystems Initiative Munich (NIM).  SD also acknowledges financial support from  the DAAD (contract No D/11/40955).}

\end{document}